\documentclass[11pt,letterpaper]{article}
\usepackage{systeme} 
\usepackage[utf8]{inputenc}
\usepackage[english]{babel}
\usepackage{amsmath}
\usepackage{amsfonts}
\usepackage{amssymb}
\usepackage{graphicx}
\usepackage[small,bf]{caption} 
\usepackage[left=3cm,right=3cm,top=2cm,bottom=3cm]{geometry}

\author{Jorge Pinochet}
\title{\textbf{Hawking for beginners: A dimensional analysis activity to perform in the classroom}}
\begin{document}

\author{Jorge Pinochet$^{*}$\\ \\
 \small{$^{*}$\textit{Departamento de Física, Universidad Metropolitana de Ciencias de la Educación,}}\\
 \small{\textit{Av. José Pedro Alessandri 774, Ñuñoa, Santiago, Chile.}}\\
 \small{e-mail: jorge.pinochet@umce.cl}\\}

\date{}
\maketitle

\begin{center}\rule{0.9\textwidth}{0.1mm} \end{center}
\begin{abstract}
\noindent In this paper we present a simple dimensional analysis exercise that allows us to derive the equation for the Hawking temperature of a black hole. The exercise is intended for high school students, and it is developed from a chapter of Stephen Hawking's bestseller \textit{A Brief History of Time}. \\ 


\begin{center}\rule{0.9\textwidth}{0.1mm} \end{center}
\end{abstract}

\maketitle

\section{Introduction}

In 1988, the shelves of the world's leading bookstores were filled with the first edition of a popular science book that was destined to become a best seller: \textit{A Brief History of Time} [1]. The author of this book, the recently deceased British physicist Stephen Hawking, had set out to captivate his readers by revealing the great mysteries of the universe without using mathematical formulas. In a chapter of his book, entitled \textit{Black Holes Ain't So Black}, Hawking explains the greatest discovery of his scientific career: the notion that black holes emit thermal radiation and have a characteristic temperature, the so-called \textit{Hawking temperature}.\\ 

Hawking's decision to omit mathematical formulas gives us an excellent pretext to turn his book into a stimulating teaching material for high school physics students. Specifically, from reading some paragraphs of \textit{Black Holes Ain't So Black} we will develop a simple dimensional analysis activity to work in classroom, which leads to the Hawking temperature equation. If the reader is a high school teacher, I hope that they will find the best way to develop the activity with his/her students and will delve into those aspects that for reasons of space we do not address in this article. The activity consists of three parts, and is intended to be developed with the guidance of the teacher.\\

In section 2 the concept of a black hole is introduced and the Hawking text that is the basis for the first part of the activity is transcribed. In section 3 the second part of the activity is presented, which consists of developing the algebra of the dimensional analysis that leads to the Hawking temperature equation. In section 4 the third and final part of the activity is developed, where a set of questions and their corresponding answers are proposed to promote better learning. The article ends with a few brief comments.

\section{First part: Black Holes Ain't So Black}

Black holes are the most extreme prediction of general relativity, which is the theory of gravity proposed by Einstein in 1915 to broaden and perfect Newton's law of gravitation. According to general relativity, a black hole is formed when a high concentration of mass, or its equivalent in energy, occurs within a closed spherical region of space called the \textit{horizon} [2]. The concentration of matter within the horizon is so high that, in order to escape its gravity, a speed greater than that of light in a vacuum is required, $c = 3\times 10^{8} m\cdot s^{-1}$. But $c$ is the maximum speed allowed by physical laws. Consequently, matter and energy can only cross the horizon from the outside to inside, but never in the opposite direction [3].\\

\begin{figure}
  \centering
    \includegraphics[width=0.4\textwidth]{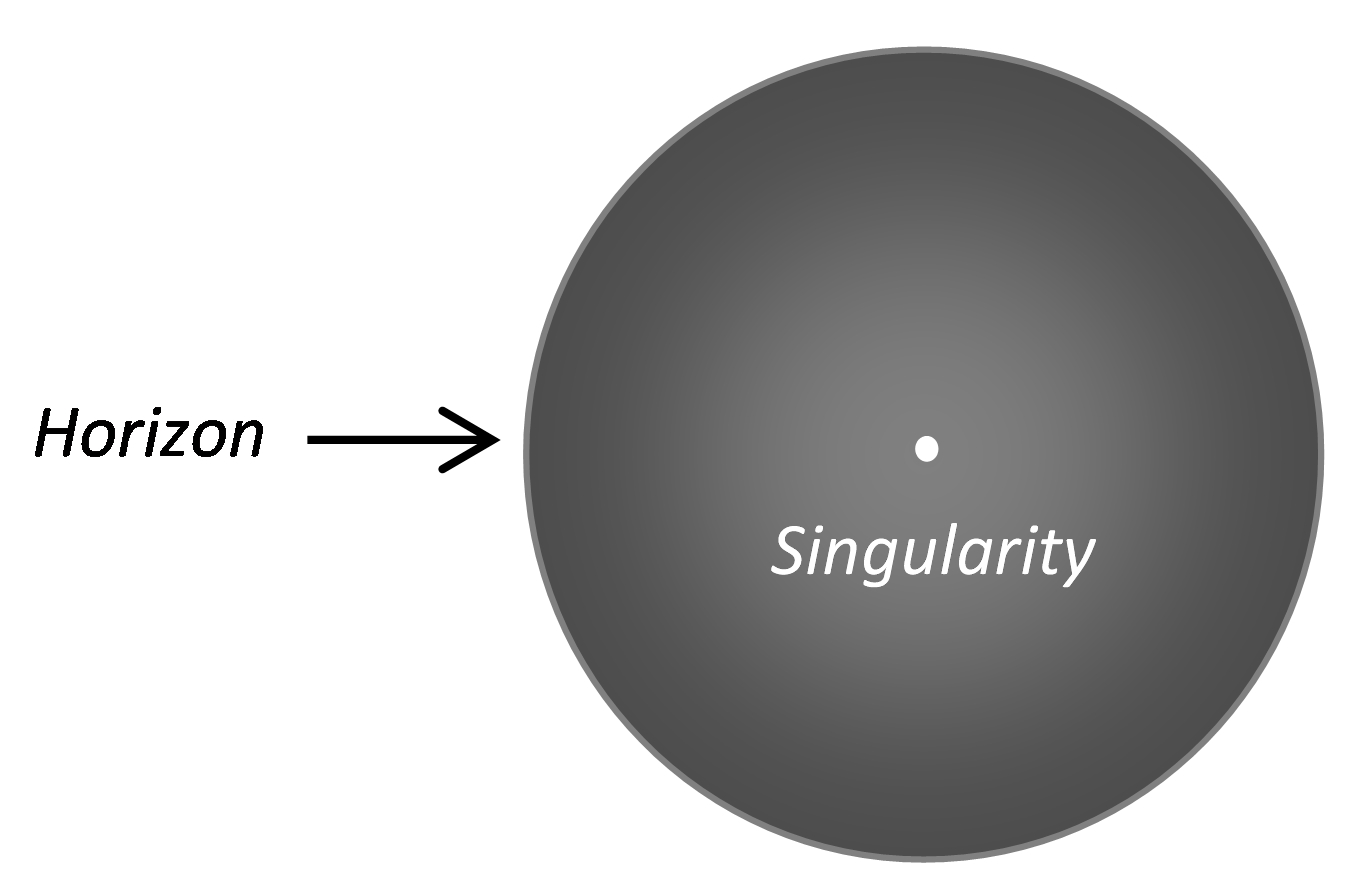}
  \caption{Within the framework of Newtonian gravitation, a black hole can be visualised as a region of space bounded by a spherical surface (horizon) of radius $R_{S}$ and at the centre is the singularity.}
\end{figure}

The entire mass of the black hole is concentrated in a mathematical point of null size and infinite density called \textit{singularity}, located in the centre of the horizon (see Fig. 1). Therefore, the horizon is not a physical surface, but it can be imagined as a unidirectional membrane that only allows the inward flow of material [1]. Within the framework of Newtonian gravity, the horizon can be intuitively visualised as a spherical surface whose radius depends only on mass\footnote{The Newtonian image is pedagogically useful but should be taken with caution. In general relativity, $R_{S}$ is only a coordinate, and does not represent the physical distance between horizon and singularity.}: 

\begin{equation}
R_{S} = \frac{2GM_{BH}}{c^{2}}.
\end{equation}

This quantity is known as \textit{gravitational radius}\footnote{The escape velocity from the surface of a massive object of radius $R$ and mass $M$ is $V_{e} = (2GM/R)^{1/2}$; taking $V_{e} = c$ and solving for $R$, we get $R = 2GM/c^{2}$. The physical meaning of this intuitive argument is the following: no form of matter or energy contained within the closed spherical surface limited by $R$ can escape, since it would need a speed greater than $c$.}, where $G = 6.67 \times 10^{-11} N\cdot m^{2} \cdot kg^{-2}$ is the universal gravitation constant, and $M_{BH}$ is the mass of the black hole, that is, the mass confined in the horizon. Considering the values of $G$ and $c$ given before and introducing the solar mass $M_{\odot} = 1.99 \times 10^{30} kg$, we can rewrite this equation in a way that allows quick calculations:

\begin{equation}
R_{S} \cong 3km \left( \frac{M_{BH}}{M_{\odot}} \right), 
\end{equation}

where $R_{S}$ is expressed in kilometres ($km$). These equations allow us to calculate the radius of the sphere within which it is necessary to compress an object so that it becomes a black hole. For example, for $M_{BH} = M_{\odot}$, $R_{S} = 3km$, which is equivalent to one hundred thousandth of the solar radius. Fig. 2 shows the graph of Eq. (2), where the horizontal axis is in units of $M_{BH}/M_{\odot}$ and the vertical axis in $km$. The graph is linear.\\

\begin{figure}
  \centering
    \includegraphics[width=0.8\textwidth]{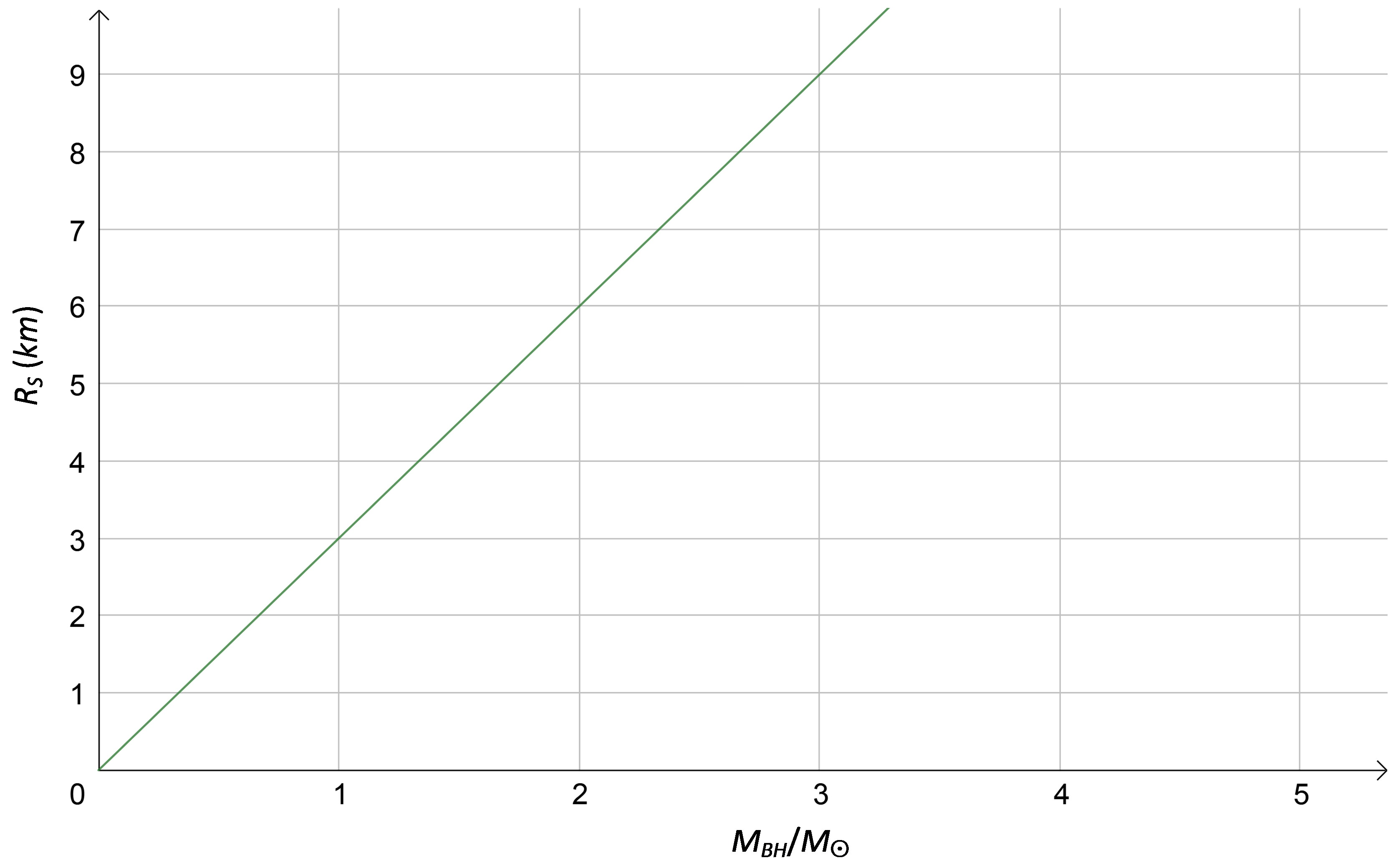}
  \caption{Gravitational radius ($R_{S}$) versus black hole mass ($M_{BH}/M_{\odot}$).}
\end{figure}

Within the framework of general relativity, a black hole can never emit radiation from its horizon, as this would imply the existence of particles that can move faster than light and cross the horizon outwards. In particular, the laws of thermodynamics ensure that the horizon temperature must be strictly null, otherwise it would emit thermal radiation and the black hole would not be black. However, Hawking discovered that it is possible to associate an absolute temperature to the horizon of a black hole, which implies that these objects emit thermal radiation in all directions. On the other hand, relativistic physics establishes that matter and energy are equivalent, so that the emitted radiation takes a part of the mass-energy of the black hole, generating a gradual reduction of $M_{BH}$. These conclusions are in flagrant conflict with general relativity and with the very definition of a black hole.\\ 

An excerpt from \textit{A Brief History of Time} is transcribed below, where Hawking explains his great discovery. The reading of this extract by students is the first part of the dimensional analysis activity.\\

In a section of \textit{Black Holes Ain't So Black} (Chapter 7, pp. 109-110), Hawking writes the following [1]:\\

\textit{(…) a black hole ought to emit particles and radiation as if it were a hotbody with a temperature that depends only on the black hole’s mass (…)}\\

At this point, Hawking goes on to explain in detail the quantum properties of black holes, as well as the concept of Hawking temperature. The British physicist's explanations end with the following words:\\

\textit{(...) A flow of negative energy into the black hole therefore reduces its mass (…) Moreover, the lower the mass of the black hole, the higher its temperature.}\\

Although we have only reproduced the beginning and the end of the Hawking text, for the proper performance of the dimensional analysis activity, it is essential that the teachers and their students read the entire text carefully.

\section{Second part: Hawking temperature through dimensional analysis}

The paragraphs from \textit{A Brief History of Time} quoted above contain all the information that is necessary for students to find the Hawking temperature $T_{H}$, which is the second part of the activity.\\

The Hawking's text reveals a key aspect: $T_{H}$ depends inversely on the mass of the black hole. Mathematically, we can express this condition as:

\begin{equation}
T_{H} \propto \frac{1}{M_{BH}}.
\end{equation}

To convert this proportionality into an equality we must find the corresponding physical constants. To achieve this, we again resort to Hawking's text, which shows us that the argument to derive $T_{H}$ depends on quantum mechanics (uncertainty principle), general relativity (gravity), and thermodynamics (temperature). This means that $T_{H}$ must be a function of the fundamental physical constants that characterise these theories: Planck constant $\hbar$ (quantum mechanics), speed of light in a vacuum $c$, gravitation constant $G$ (general relativity), and Boltzmann constant $k_{B}$ (thermodynamics)\footnote{Planck constant, $\hbar = 1.05 \times 10^{-34} J\cdot s$, plays a central role in the quantum mechanics and is named after its discoverer, the German physicist Max Planck, a founding father of that theory. The speed of light in vacuum, $c = 3\times 10^{8} m\cdot s^{-1}$, is a universal constant postulated by Albert Einstein as the foundation of his theory of special relativity. The gravitational constant, $G = 6.67 \times 10^{-11} N\cdot m^{2} \cdot kg^{-2}$, determines the intensity of the gravitational pull force between bodies, and appears both in Newton's law of gravitation and in Einstein's general theory of relativity. The Boltzmann constant, $k_{B} = 1.38 \times 10^{-23} J\cdot K^{-1}$, is named after the Austrian physicist Ludwig Boltzmann, who made important contributions to the theory of statistical mechanics, in whose fundamental equations kB plays a central role.} . We can express this condition as: 

\begin{equation}
T_{H} = \frac{\alpha \beta}{M_{BH}},
\end{equation}

thus: 

\begin{equation}
M_{BH} T_{H} = \alpha \beta,
\end{equation}

where $\alpha = \alpha (\hbar, c, G, k_{B})$ is a quantity that is expressed as a combination of the indicated physical constants, and $\beta$ is a pure number, such as $4$, $\sqrt{3}$, or $2\pi$. We must include $\beta$ because is a dimensionless quantity, and therefore dimensional analysis cannot determine its value. Our goal will be to find the value of $\alpha$. For this, it will be necessary to algebraically combine different powers of $\hbar$, $c$, $G$, and $k_{B}$ so that the relation (5) is dimensionally satisfied. Let's start by writing this relation using the standard notation of dimensional analysis:

\begin{equation}
\left[ T_{H} \right] \left[ M_{BH} \right] = \left[ k_{B} \right]^{p} \left[ \hbar \right]^{q} \left[ c \right]^{r} \left[ G \right]^{s},
\end{equation}

where $\left[ X \right]$ is the conventional symbol to designate "the dimension of $X$". Since $\beta$ is a dimensionless quantity, it must have $\left[ \beta \right] = 1$. This is the reason why $\beta$ does not appear in Eq. (6). This equation reveals that determining the value of $\alpha$ is equivalent to finding the values of $p$, $q$, $r$, and $s$. Let us start by introducing the usual symbols of the dimensional analysis, with $M$ for mass, $L$ for length, $T$ for time, and $\Theta$ for absolute temperature. For the right side of Eq. (6) we must have:

\begin{equation}
\left[ T_{H} \right] = \Theta = \Theta^{1},
\end{equation}

\begin{equation}
\left[ M_{BH} \right] = M = M^{1}.
\end{equation}

As we know the units in which the constants $k_{B}$, $\hbar$, $c$, and $G$ are measured, for the left side of Eq. (6) we obtain:

\begin{equation}
\left[ k_{B} \right] = M^{1} L^{2} T^{-2} \Theta ^{-1},
\end{equation}

\begin{equation}
\left[ \hbar \right] = M^{1} L^{2} T^{-1},
\end{equation}

\begin{equation}
\left[ c \right] = L^{1} T^{-1},
\end{equation}

\begin{equation}
\left[ G \right] = M^{-1} L^{3} T^{-2}.
\end{equation}

Replacing Eqs. (7), (8), (9), (10), (11), and (12) in Eq. (6) we get:

\begin{equation}
\Theta^{1} M^{1} = (M^{1} L^{2} T^{-2} \Theta ^{-1})^{p} (M^{1} L^{2} T^{-1})^{q} (L^{1} T^{-1})^{r} (M^{-1} L^{3} T^{-2})^{s}.
\end{equation}

To make it clear that the left side is independent of $L$ and $T$ it is necessary to introduce these two dimensions raised to a zero exponent:

\begin{equation}
L^{0} T^{0} \Theta^{1} M^{1} = (M^{1} L^{2} T^{-2} \Theta ^{-1})^{p} (M^{1} L^{2} T^{-1})^{q} (L^{1} T^{-1})^{r} (M^{-1} L^{3} T^{-2})^{s}.
\end{equation}

Ordering the terms:

\begin{equation}
L^{0} T^{0} \Theta^{1} M^{1} = L^{2p+2q+r+3s} T^{-2p-q-r-2s} \Theta^{-p} M^{p+q-s}.
\end{equation}

For this equality to be satisfied, the exponent of each dimension on the left side must be equal to the exponent of the corresponding dimension on the right side. By proceeding in this way, the following system of four equations with four unknowns is obtained:

\begin{equation}
  \systeme{
  0 = 2p + 2q + r + 3s, 
  0 = -2p - q - r - 2s, 
  1 = -p, 
  1 = p + q - s 
  }.
\end{equation}

It is observed that the number of equations is determined by the number of dimensional parameters ($L$, $T$, $\Theta$, $M$), while the number of unknowns is determined by the number of variables ($k_{B}$, $\hbar$, $c$, $G$). The system of equations can be simplified if we enter the value $p = -1$ (third equation) in the remaining equations:

\begin{equation}
  \systeme{
  2 = 2q + r + 3s, 
  2 = q+r+2s, 
  2 = q-s 
  }.
\end{equation}

Since the number of equations is equal to the number of unknowns, this system has a unique solution. The simplest procedure to solve this system is to apply Cramer's rule, the description of which can be found in most of the introductory texts for algebra [4, 5]. After applying this rule, it is found that $q = 1$, $r = 3$, and $s = –1$. It can be verified by simple substitution that these values satisfy the equation system (17). Entering the values of $p$, $q$, $r$, and $s$ in Eq. (6):

\begin{equation}
\left[ T_{H} \right] \left[ M_{BH} \right] = \left[ k_{B} \right]^{-1} \left[ \hbar \right]^{1} \left[ c \right]^{3} \left[ G \right]^{-1}.
\end{equation}

If we eliminate the dimensional brackets on both sides of this equality, it is necessary to restore the dimensionless constant $\beta$:

\begin{equation}
T_{H} M_{BH} = \beta k_{B}^{-1} \hbar c^{3} G^{-1}.
\end{equation}

Finally we get:

\begin{equation}
T_{H} = \beta \frac{\hbar c^{3}}{k_{B}G M_{BH}}.
\end{equation}

\begin{figure}
  \centering
    \includegraphics[width=0.8\textwidth]{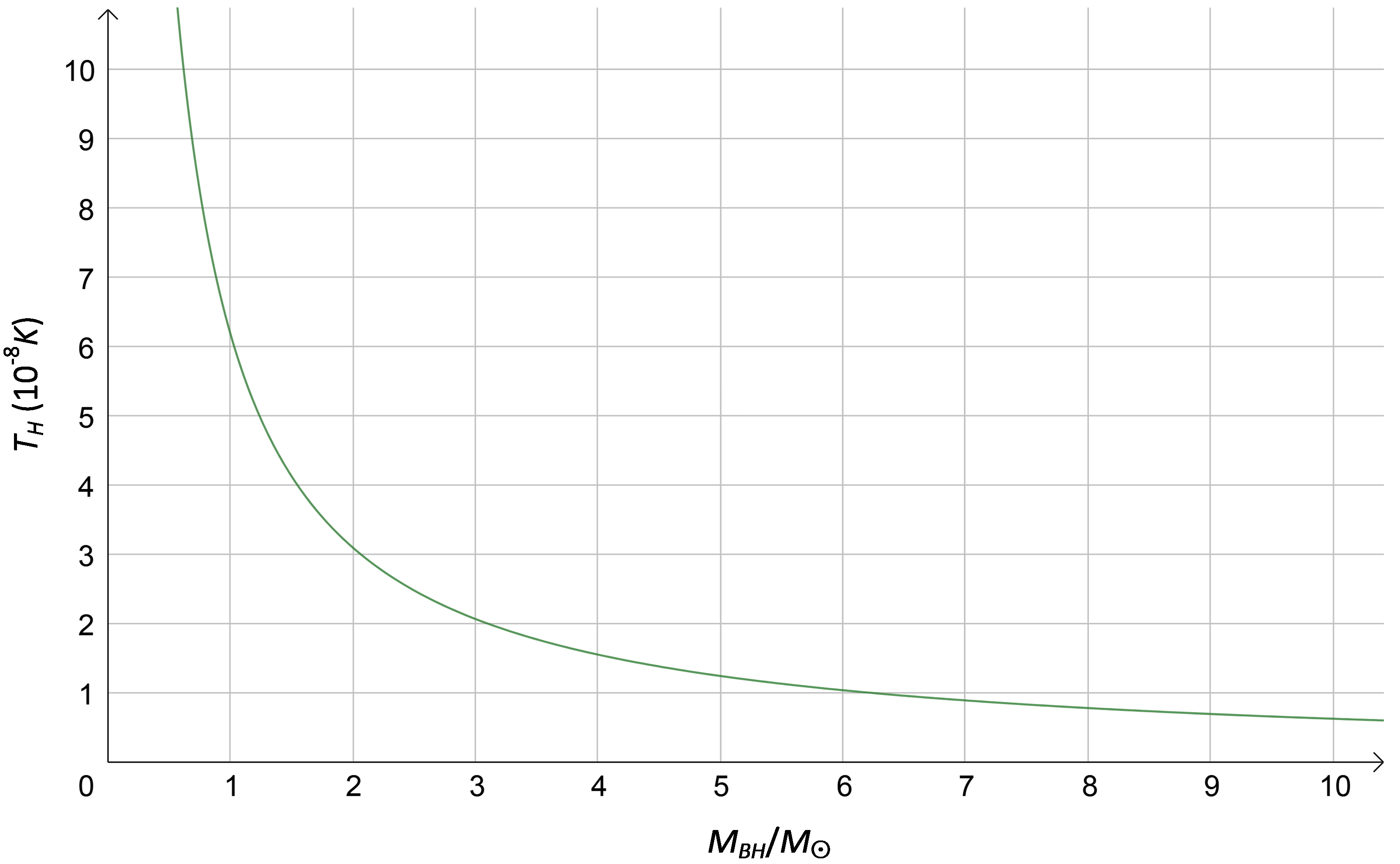}
  \caption{Hawking temperature ($T_{H}$) versus black hole mass ($M_{BH}/M_{\odot}$).}
\end{figure}

The exact expression for $T_{H}$ originally found by Hawking is\footnote{It should be clarified that this is not the temperature on the horizon; it is the temperature that an observer located at a great distance, ideally infinite, would measure.} [2, 6-8]: 

\begin{equation}
T_{H} = \frac{\hbar c^{3}}{8\pi k_{B}G M_{BH}}.
\end{equation}

Then, except for the dimensionless constant $\beta$, we have reached the same equation found by Hawking. Comparing Eqs. (20) and (21) we observe that $\beta = 1/8\pi$. Introducing the solar mass $M_{\odot} = 1.99 \times 10^{30} kg$ and considering that $\hbar = h/2\pi = 1.05 \times 10^{-34} J\cdot s$  and $k_{B} = 1.38 \times 10^{-23} J\cdot K^{-1}$ we can write the Hawking equation in a form that allows quick calculations:

\begin{equation}
T_{H} = 6.17 \times 10^{-8} K \left( \frac{M_{\odot}}{M_{BH}} \right),
\end{equation}

where $T_{H}$ is expressed in Kelvin ($K$). Fig. 3 shows the graph of Eq. (22), where the horizontal axis is in units of $M_{BH}/M_{\odot}$ and the vertical axis in units of $10^{-8} K$. For example, for $M_{BH} = 1M\odot$, $T_{H} = 6.17 \times 10^{-8} K$. It is observed that $T_{H}$ asymptotically approaches zero as $M_{BH}$ grows. In the following section the physical meaning of Eqs. (21) and (22) is analysed in more detail.

\section{Third part: Questions and problems}

The following questions are the last part of the activity. The purpose of the questions is that students gain a deeper understanding of Eq. (21) and of the concept of the Hawking temperature. \\ 

\textbf{1}. In order for the Hawking temperature to be detected by astronomical observations, must the mass $M_{BH}$ of a black hole be small or large? \\  
\textbf{R}: The smaller $M_{BH}$ is, the greater $T_{H}$ is and the easier the detection.\\ 

\textbf{2}. The least massive black holes for which there is observational evidence are the \textit{stellar black holes}, whose masses are of the order of the solar mass, $M_{\odot} = 1.99 \times 10^{30}kg$. What is the Hawking temperature of these objects? Is it possible to detect it?\\ 
\textbf{R}: $T_{H} \sim 10^{-7} K$, and is too low to be detected. \\ 

\textbf{3}. Hawking speculated on the possible existence of so-called \textit{micro black holes}, whose masses are of the order of $10^{12} kg$ or less (this is the mass of an average asteroid or a mountain). Are these objects detectable?\\ 
\textbf{R}: $T_{H} \sim 10^{11} K$. As this temperature is very high, in principle it is detectable.\\ 

\textbf{4}. What gravitational radius would a micro black hole have? Find out the radius of a proton (or a neutron) and compare it with the value you obtained for a micro black hole.\\ 
\textbf{R}: $R_{S} \sim 10^{-15} m$, and it is equivalent to the radius of a proton.\\ 

\textbf{5}. The \textit{Wien displacement law} relates the maximum wavelength $\lambda$ ($m$) of a hot body with its absolute temperature $T$ ($K$) by the equation $\lambda T = 2.9 \times 10^{-3} m\cdot K$. Determine $\lambda$ for the Hawking temperature of a stellar black hole and a micro hole. Find out which domain of the electromagnetic spectrum the calculated values correspond to.\\ 
\textbf{R}: $\lambda_{stellar} \sim 10^{4} m$, radio waves; $\lambda_{micro} \sim 10^{-14} m$, gamma rays. \\ 

\textbf{6}. Specialists agree that Hawking's findings about black holes are among the most important contributions to physics in recent decades. However, Hawking did not receive the Nobel Prize. From the information given before, and based on the results you obtained, how do you explain this?\\ 
\textbf{R}: The award of the Nobel Prize in Physics requires that the theories be supported by empirical evidence, but as is mainly deduced from problems 1 and 2, no solid evidence has yet been found in favour of Hawking's discoveries.

\section{Final comments}

This activity reveals the power and usefulness of dimensional analysis. However, we have seen that this procedure does not deliver the values of dimensionless constants, but this does not prevent it from being a widely used tool that has led to important discoveries in physics. In fact, despite the extraordinary mathematical complexity of the reasoning originally used by Hawking to derive Eq. (21), dimensional analysis has allowed us to approach this equation using high school algebra. Obviously, without the guidance of the Hawking text transcribed in Section 2 this would not have been possible. \\ 

In any case, I hope that the activity presented arouses students' interest in physics and motivates them to continue learning about Hawking's great contributions to our current understanding of black holes.

\section*{Acknowledgments}
I would like to thank to Daniela Balieiro for their valuable comments in the writing of this paper. 

\section*{References}

[1] S.W. Hawking, A brief history of time, Bantam Books, New York, 1998.

\vspace{2mm}

[2] J. Pinochet, The Hawking temperature, the uncertainty principle and quantum black holes, Phys. Educ., 53. (2018) 1-6

\vspace{2mm}

[3] J.P. Luminet, Black Holes, Cambridge University Press, Cambridge, 1999.

\vspace{2mm}

[4] D.C. Lay, Álgebra lineal y sus aplicaciones, 3 ed., Pearson Educación, México, D.F., 2007.

\vspace{2mm}

[5] D. Poole, Álgebra lineal: Una introducción moderna, 3 ed., Cengage Learning, México, D.F., 2011.

\vspace{2mm}

[6] S.W. Hawking, Particle creation by black holes, Communications in Mathematical Physics, 43 (1975) 199-220.

\vspace{2mm}

[7] S.W. Hawking, Black Hole explosions?, Nature, 248 (1974) 30-31.

\end{document}